\documentclass[12pt]{article}

\usepackage{amssymb,latexsym,amsmath}

\begin{document}

\sloppy
\renewcommand{\theequation}{\arabic{section}.\arabic{equation}}
\thinmuskip = 0.5\thinmuskip
\medmuskip = 0.5\medmuskip
\thickmuskip = 0.5\thickmuskip
\arraycolsep = 0.3\arraycolsep

\newtheorem{theorem}{Theorem}[section]
\newtheorem{lemma}[theorem]{Lemma}
\renewcommand{\thetheorem}{\arabic{section}.\arabic{theorem}}

\newcommand{\prf}{\noindent{\bf Proof.}\ }
\def\prfe{\hspace*{\fill} $\Box$

\smallskip \noindent}

\def\be{\begin{equation}}
\def\ee{\end{equation}}
\def\bea{\begin{eqnarray}}
\def\eea{\end{eqnarray}}
\def\beas{\begin{eqnarray*}}
\def\eeas{\end{eqnarray*}}

\newcommand{\R}{\mathbb R} 
\newcommand{\N}{\mathbb N}

\def\supp{\mathrm{supp}\,} 
\def\vol{\mathrm{vol}\,} 
\def\sign{\mathrm{sign}\,}
\def\ekin{E_\mathrm{kin}}
\def\epot{E_\mathrm{pot}}

\def\C{{\cal C}}
\def\H{{\cal H}}
\def\Hc{{{\cal H}_C}}

\title{A non-variational approach to nonlinear stability in stellar 
       dynamics applied to the King model}
\author{ Yan Guo\\
         Lefschetz Center for Dynamical Systems \\
         Division of Applied Mathematics \\
         Brown University, Providence, RI 02912 \\
         and \\
         Gerhard Rein\\
         Mathematisches Institut der
         Universit\"at Bayreuth\\
         D 95440 Bayreuth, Germany}
\maketitle

\begin{abstract}
In previous work by Y.~Guo and G.~Rein, nonlinear stability of equilibria
in stellar dynamics, i.e., of steady states of the Vlasov-Poisson system,
was accessed by variational techniques. Here we propose a different,
non-variational technique and use it to prove nonlinear stability 
of the King model against a class of spherically symmetric, 
dynamically accessible perturbations.
This model is very important in astrophysics and
was out of reach of the previous techniques.

\end{abstract}

\section{Introduction}
\setcounter{equation}{0}

In astrophysics a galaxy or a globular cluster is often modeled as a large ensemble 
of particles, i.e., stars, which interact only by the gravitational field which they
create collectively, collisions among the stars being sufficiently rare to
be neglected. The time evolution of the distribution function $f=f(t,x,v)\geq 0$ 
of the stars in phase space is then given by the Vlasov-Poisson system:
\be
\partial_{t}f+v\cdot \nabla_{x}f-\nabla_{x}U\cdot \nabla_{v}f =0,
\label{vlasov}
\ee
\be
\Delta U = 4\pi \rho,\ \lim_{|x| \to \infty} U(t,x) = 0, \label{poisson}
\ee
\be
\rho(t,x) = \int f(t,x,v)dv \label{rhodef}.  
\ee
Here $t\in\R$ denotes time, and $x,v\in \R^3$ denote position and velocity,
$U=U(t,x)$ is the gravitational potential of the ensemble, and $\rho=\rho(t,x)$
is its spatial density.

A well-known approach to obtaining steady states of the Vlasov-Poisson system
is to make an ansatz of the form
\[
f_{0}(x,v) = \phi(E),
\]
where the particle energy $E$ is defined as
\[
E:=\frac{1}{2}|v|^2+U_{0}(x).
\]
Since for a time-independent potential $U_0=U_{0}(x)$ the particle energy
is conserved along the characteristics of the Vlasov equation (\ref{vlasov})
it remains to make sure that for the chosen $\phi$ the resulting semilinear
Poisson equation
\be \label{semilinpoisson}
\Delta U_0 =  4\pi \int \phi \left(\frac{1}{2}|v|^2+U_{0}\right)\, dv
\ee
has a solution. For a very large class of ansatz functions $\phi$
this approach leads to steady states with finite mass and compact support,
cf.\ \cite{RR}. Steady states where as above
the distribution of the particles in phase space
depends only on the particle energy are usually called 
{\em isotropic} and are always spherically symmetric.
A central questions in stellar dynamics, which has attracted considerable attention in the
astrophysics literature, cf.\ \cite{BT,FP} and the references there, is the 
dynamical stability of such steady states. 

The Vlasov-Poisson system is conservative, i.e., the total energy
\[
\H (f) := \ekin (f) + \epot (f)
= \frac{1}{2} \int |v|^2 f(x,v)\,dv\,dx - \frac{1}{8 \pi} \int |\nabla U_f(x)|^2 dx
\]
of a state $f$ is conserved along solutions and hence is a natural candidate for a 
Lyapunov function in a stability analysis; $U_f$ denotes the potential
induced by $f$. However, the energy does not have critical
points, i.e., the linear part in an expansion about
any state $f_0$ with potential $U_0$ does not vanish:
\[
\H(f) = \H(f_0) + 
\int\!\!\!\!\int \left(\frac{1}{2} |v|^2 + U_0 \right)(f-f_0)  \,dv\,dx
- \frac{1}{8 \pi} \int|\nabla U_f-\nabla U_0 |^2 dx .
\]
To remedy this situation one observes that 
for any reasonable function $\Phi$ the so-called {\em Casimir functional}
\[
\C(f) := \int\!\!\!\!\int \Phi(f(x,v))\,dv\,dx
\]
is conserved as well. If the energy-Casimir functional 
\[
\Hc := \H + \C
\]
is expanded about an isotropic steady state, then
\beas
\Hc (f) 
&=& 
\Hc (f_0) + 
\int\!\!\!\!\int (E + \Phi'(f_0))\,(f-f_0)  \,dv\,dx \nonumber \\
&& 
{}- \frac{1}{8 \pi} \int|\nabla U_f-\nabla U_0 |^2 dx 
+ \frac{1}{2} \int\!\!\!\!\int \Phi''(f_0) (f-f_0)^2 \,dv\,dx 
+ \ldots \\
&=:&
\Hc (f_0) + D \Hc(f_0)[f-f_0] + D^2 \Hc(f_0)[f-f_0] + \ldots . 
\eeas
Now one can try to choose $\Phi$ is such a way that at least formally
$f_0$ is a critical point of the energy-Casimir functional,
i.e., $\Phi^{\prime }(f_{0})= -E$. If $\phi' < 0$ on the support 
of $f_0$ the former relation holds on this support if $\Phi' = -\phi^{-1}$.
The formal second order variation in the above expansion then takes the form 
\be \label{h2}
D^2\Hc (f_0)[g] =  \frac{1}{2}\int\!\!\!\!\int_{\{f_{0}>0\}}
\frac{1}{-\phi'(E)} g^2\, dv\, dx 
-\frac{1}{8\pi}\int |\nabla_{x}U_{g}|^2 dx.  
\ee
It is natural to expect that positive definiteness of this quadratic 
form should 
imply stability for $f_{0}$. Ever since the seminal work of {\sc Antonov} \cite{An}
there have been vigorous efforts in the astrophysics community  to establish
this positive definiteness and to derive stability results in this way.

An important step in this direction was to show that the above
quadratic form is positive definite on linearized, dynamically accessible
perturbations. To make this precise
we define the Lie-Poisson bracket of two functions $f_1, f_2$ 
of $x$ and $v$ as 
\begin{equation}
\{f_{1},f_{2}\} := \nabla _{x}f_{1}\cdot \nabla _{v}f_{2}-\nabla
_{v}f_{1}\cdot \nabla _{x}f_{2}.  \label{lie}
\end{equation}
Then the following holds:
\begin{lemma} \label{ks}
Let $\phi^{\prime }<0$ and let $h\in
C_{c}^{\infty }(\R^6)$ be spherically symmetric with support in the set 
$\{f_{0}>0\}$  and such that $h(x,-v)=-h(x,v)$. Then 
\[
D^2\Hc (f_0)[\{f_{0},h\}]
\geq 
-\frac{1}{2} \int_{f_{0}>0} \phi^{\prime }(E)
\left[ |x\cdot v|^2
\left\vert \left\{ E,\frac{h}{x\cdot v}\right\} \right\vert^2
+ \frac{1}{r}U_{0}^{\prime}\, h^2
\right] \,dv\,dx.
\]
\end{lemma}
Here $U_0'$ denotes the radial derivative of the steady state potential.
Since $U_0$ is radially increasing, the right hand side in the estimate above
is indeed positive for $h\neq 0$. We refer to \cite{KS,SDLP} for astrophysical
investigations where this result is used to analyze
linearized stability. We do not go into the reasons why perturbations of
the form $\{f_0,h\}$ are called dynamically accessible for the linearized system,
but for the sake of completeness we provide a proof of this elegant result in
the Appendix. Despite its significance the result is still quite a distance 
away from a true, nonlinear stability result. 
There are at least two serious mathematical
difficulties. Firstly, it is very challenging to use the positivity of 
$D^2 \Hc (f_{0})[g]$ to control the higher order remainder in the expansion
of the energy-Casimir functional \cite{Wa}. This is due to the
non-smooth nature of $f_{0}=\phi(E)$ in all important examples. Secondly,
even if one succeeds in controlling the higher order terms, the positivity 
of $D^2 \Hc (f_{0})[g]$ in the lemma is only valid for certain
perturbation of the form $g=\{f_{0},h\}$. This class of perturbations is
invariant under solutions of the linearized Vlasov-Poisson system,
but it is not invariant under solutions of the nonlinear system.

To overcome these difficulties a variational approach was
initiated by {\sc Wolansky} \cite{Wo1} and then developed 
systematically by {\sc Guo}
and {\sc Rein} \cite{G1,G2,GR1,GR2,GR3,GR4,R1,R2,RG}. Their 
method entirely avoids the delicate
analysis of the second order term $D^2 \Hc (f_{0})$ in 
(\ref{h2}), and it has led to the first rigorous nonlinear stability 
proofs for a large class of steady states. More precisely, a large class 
of steady states is obtained as minimizers of energy-Casimir functionals
under a mass constraint $\int f = M$,
and their minimizing property then entails their stability.
In particular, all polytropes $f_{0}(x,v)=(E_{0}-E)_{+}^{k}$ 
with $0<k\leq 7/2$ are covered; here $E_0<0$ is a certain cut-off 
energy, and $(\cdot)_+$ denotes the positive part. 
For $k >7/2$ the corresponding steady state has infinite mass
and is therefore unphysical. In addition, many new stable galaxy models were
established. The variational method has also been investigated in 
\cite{DSS,H,LMR,SS,Wo2}.

Despite its considerable success, the variational approach has drawbacks
and limitations, the main one being that by its very nature it can not
access the stability of steady states which are only local, but not global
minimizers of the energy-Casimir functional.
Since the existence of the steady state as a (global) minimizer
is aimed for, certain growth conditions on the Casimir function $\Phi$ are needed,
which are not satisfied for all steady states with $\phi' <0$. 
Most notably, the King model obtained by
\[
f_0(x,v) = (e^{E_0 - E}-1)_+
\]
is the single most important model which is currently out of the reach. 
It describes isothermal galaxies and is widely used in astrophysics.
The corresponding Casimir function (\ref{q}) 
has very slow growth for $f \to \infty$, and as a result the
variational method fails. 

The aim of the present paper is to develop a new approach
to nonlinear stability
results for steady states which need not be global minimizers of the 
corresponding energy-Casimir functional by exploiting Lemma~\ref{ks}.
Although we are aiming for a general approach,
we focus here on the King model and as a first step 
establish its nonlinear stability against
spherically symmetric, dynamically accessible perturbations.

The paper proceeds as follows. In the next section we formulate our results.
The nonlinear stability of the King model is an easy corollary of
the following main theorem: In a certain neighborhood of the King model
the potential energy distance of a perturbation can be controlled
in terms of the energy-Casimir distance.
In particular, within a certain class of perturbations, which
is invariant under solutions of the nonlinear Vlasov-Poisson system,
the King model is a local minimizer of the corresponding energy-Casimir
functional. 
The resulting stability estimate is more explicit that the
ones obtained by the variational approach. The main part of the work
is then done in Section~3 where the local minimizing property
of the King model is established. In an appendix we give a proof of 
Lemma~\ref{ks}.

\section{Main results}
\setcounter{equation}{0} 

We start with a steady state $f_0$ with induced potential
$U_0$ and spatial density $\rho_0$, satisfying the relation
\begin{equation}
f_{0}(x,v)=\phi_0(E) := \left( e^{E_{0}-E}-1\right) _{+},\ E:=\frac{1}{2}|v|^2+U_{0}(x).  \label{king}
\end{equation}
The cut-off energy $E_{0}<0$ is a given negative constant and 
$(\cdot)_+$ denotes the positive part. 
The corresponding Casimir function in the sense of the introduction is
\begin{equation}
\Phi_{0}(f) := (1+f)\ln (1+f)-f.  \label{q}
\end{equation}
The existence of such King models,
i.e., of suitable solutions of the resulting semilinear Poisson equation
(\ref{semilinpoisson}), is established in \cite{RR}. Such a  model has
compact support
\[
\supp f_0 = \{(v,v) \in \R^6 \mid E(x,v) \leq E_0\} =: \{E \leq E_0\},
\]
and it is spherically symmetric. A state $f$ is called 
{\em spherically symmetric}
if for any rotation $A\in \mathrm{SO}(3)$, 
\[
f(x,v) = f(A x,A v),\ x,v \in \R^3. 
\] 
It is well known that non-negative, smooth, and compactly supported
initial data $f(0) \in C^1_c(\R^6)$ launch unique global smooth solutions
$t\mapsto f(t)$
of the Vlasov-Poisson system \cite{Pf,LP,Sch}. 
If the initial datum is spherically symmetric then this symmetry is preserved,
and the modulus of the particle angular momentum squared,
\[
L:= |x \times v|^2 = |x|^2 |v|^2 - (x\cdot v)^2 ,
\]
is conserved along characteristics of the Vlasov equation.
Hence for any smooth function $\Phi$ such that $\Phi(0,L)=0,\ L\geq 0$,
the functional 
\[
\int\!\!\!\!\int \Phi(f,L)\,dv\,dx
\]
is conserved along spherically symmetric solutions of the 
Vlasov-Poisson system;
unless explicitly stated otherwise integrals $\int$ always extend over $\R^3$.
We consider the following class of perturbations:
\beas
\mathcal{S}_{f_{0}} :=  \Bigl\{f\in L^1(\R^6) 
&\mid& 
f\geq 0 \ \mbox{spherically symmetric}, \\
&& \int\!\!\!\!\int\Phi(f,L)=\int\!\!\!\!\int \Phi(f_{0},L) 
\text{ for all }\Phi\in C^2([0,\infty[^2) \text{ with } \\
&&
\Phi(0,L)=\partial_f \Phi(0,L)= 0,\ L\geq 0,\ \mbox{and}\ 
\partial_f^2 \Phi \ \mbox{bounded}\Bigr\}.
\eeas
As noted above, the class $\mathcal{S}_{f_{0}} \cap C^1_c(\R^6)$ is
invariant under solutions of the Vlasov-Poisson system.
Moreover, functions in $\mathcal{S}_{f_{0}}$ are equi-measurable to
$f_0$, i.e., for every $\tau>0$ the sets $\{ f>\tau\}$ and $\{ f_0>\tau\}$ have 
the same measure,
in particular, $||f||_p = ||f_0||_p$ for any $L^p$-norm, $p\in [1,\infty]$. 

For the Casimir function $\Phi_0$ defined in (\ref{q})
we define the energy-Casimir functional
as in the introduction.
Then for $f\in \mathcal{S}_{f_{0}}$ we have
\begin{eqnarray*}
\Hc(f)-\Hc(f_{0}) 
&=&
\int\!\!\!\!\int [\Phi_{0}(f)-\Phi_{0}(f_{0})+(E-E_{0})(f-f_{0})]\,dv\,dx \\
&&-\frac{1}{8\pi }\int |\nabla U_{f}-\nabla U_0|^2 dx;
\end{eqnarray*}
notice that $\int f = \int f_0$ which allows us to bring 
in the term $E_0(f-f_0)$. Now
\[
(E-E_0) (f-f_0) \geq -\Phi_0'(f_0) (f-f_0)
\]
with equality on the support of $f_0$, and hence for 
$f\in \mathcal{S}_{f_{0}}$,
\bea \label{convest}
\Phi_{0}(f)-\Phi_{0}(f_{0})+(E-E_{0})(f-f_{0}) 
&\geq&
\frac{1}{2} \inf_{0\leq \tau \leq ||f_0||_\infty} \Phi_0''(\tau)\; (f-f_0)^2 \nonumber \\
&\geq& 
C_0 (f-f_0)^2
\eea
where $C_0:= 1/(2+ 2 ||f_0||_\infty)$; notice again that 
$||f||_\infty = ||f_0||_\infty$ for any 
$f\in \mathcal{S}_{f_{0}}$.  
The deviation from the steady state is going to be measured
by the quantity
\bea \label{d}
d(f,f_{0})
&:=&
\int\!\!\!\!\int 
[\Phi_{0}(f)-\Phi_{0}(f_{0})+(E-E_{0})(f-f_{0})]\,dv\,dx \nonumber \\
&&
{}+\frac{1}{8\pi }
\int |\nabla U_{f}-\nabla U_0|^2 dx,
\eea
which, as we have seen, controls both $||f-f_0||_2$ and 
$||\nabla U_{f}-\nabla U_0||_2$,
and satisfies the following relation to the energy-Casimir functional:
\be \label{decrel}
d(f,f_{0}) = \Hc(f)-\Hc(f_{0}) + 
\frac{1}{4\pi }\int |\nabla U_{f}-\nabla U_0|^2 dx.
\ee
Our stability result is the following:
\begin{theorem} \label{main}
There exist constants $\delta >0$ and $C>0$ such that 
for any solution $t\mapsto f(t)$ of the Vlasov-Poisson system
with $f(0) \in \mathcal{S}_{f_{0}}\cap C_{c}^{1}(\R^6)$ and 
\[
d(f(0),f_{0}) \leq \delta 
\]
the estimate
\[
d(f(t),f_{0}) \leq C \; d(f(0),f_{0})
\]
holds for all time $t>0$.
\end{theorem}
\noindent
{\bf Remark:}
In order to better understand the perturbation class $\mathcal{S}_{f_{0}}$
we show that it contains 
{\em spherically symmetric, dynamically accessible
perturbations} 
by which we mean the following:
Let $H=H(x,v) \in C^2(\R^6)$ be spherically symmetric,
and let $g=g(s,x,v)$ denote the solution of the linear problem
\[
\partial_s g + \nabla_v H\cdot \nabla_x g - \nabla_x H\cdot \nabla_v g = 0,
\quad \mbox{i.e.},\quad
\partial_s g(s) = \{H,g(s)\},
\]
with initial datum $g(0) = f_0$; we assume that $H$ is such that 
this solution exists on some interval $I$ about $s=0$. Then for any $s \in I$,
$g(s)$ is spherically symmetric and equi-measurable with $f_0$.
Moreover, for any function $\Phi$ as considered in the definition
of  $\mathcal{S}_{f_{0}}$,
\[
\partial_s \Phi(g(s),L) = \partial_f \Phi(g(s),L) \{ H,g(s)\},
\]
and by a simple computation, $\{H,L\} = 0$.
Hence
\beas
\frac{d}{ds} \int\!\!\!\!\int \Phi(g(s),L)\,dv\,dx
&=& 
\int\!\!\!\!\int 
\left( \partial_f \Phi \{H,g(s)\} +  \partial_L \Phi \{H,L\}\right)\,dv\,dx \\
&=&
\int\!\!\!\!\int \{H,\Phi(g(s),L)\}\,dv\,dx = 0
\eeas
after an integration by parts.
This means that $g(s) \in \mathcal{S}_{f_{0}}$ for any such 
generating function $H$ and any $s$. The only undesirable restriction 
in the class $\mathcal{S}_{f_{0}}$
or in the generating functions $H$ respectively is the spherical symmetry
which hopefully can be removed in the future.

\smallskip

The stability result Theorem~\ref{main} is easily deduced from the
following theorem:
\begin{theorem} \label{lower}
There exist constants $\delta _{0}>0,$ and $C_0>0$ such that
for all $f\in \mathcal{S}_{f_{0}}$ with $d(f,f_{0})\leq \delta _{0}$ 
the following estimate holds:
\[
\Hc(f)-\Hc(f_{0})\geq C_0 ||\nabla U_{f}-\nabla U_0||_2^2.  
\]
\end{theorem}
Before going into the proof of this theorem, which will occupy the rest of this
paper, we conclude this section by deducing our stability result from it.

\noindent
{\bf Proof of Theorem~\ref{main}}. 
Let $\delta:= \delta_0 (1+1/(4 \pi C_0))^{-1}$
with $\delta_0$ and $C_0$ from Theorem~\ref{lower}. Consider a solution
$t\mapsto f(t)$ of the Vlasov-Poisson system with 
$f(0) \in \mathcal{S}_{f_{0}}\cap C_{c}^{1}(\R^6)$ and 
\[
d(f(0),f_{0}) \leq \delta < \delta_0. 
\]
Then by continuity we can choose some maximal $t^\ast \in ]0,\infty]$ such that
\[
d(f(t),f_{0}) < \delta_0,\ t\in [0,t^\ast[. 
\]
Now $f(t) \in \mathcal{S}_{f_{0}}$ for all $t$, and hence
Theorem~\ref{lower}, the relation (\ref{decrel}) of $d$ to the energy-Casimir 
functional, and the fact that the latter is a conserved quantity
yield the following chain of estimates for $t\in[0,t^\ast[$:
\beas
d(f(t),f_0)
&=&
\Hc(f(t)) - \Hc(f_0) + \frac{1}{4\pi} ||\nabla U_{f(t)}-\nabla U_0||_2^2\\
&\leq&
\Hc(f(t)) - \Hc(f_0) + \frac{1}{4 \pi C_0} \left(\Hc(f(t)) - \Hc(f_0)\right)\\
&=&
\left(1+\frac{1}{4 \pi C_0}\right) \, \left(\Hc(f(0)) - \Hc(f_0)\right)\\
&\leq&
\left(1+\frac{1}{4 \pi C_0}\right)\,d(f(0),f_0) < \delta_0.
\eeas
This implies that $t^\ast = \infty$, and Theorem~\ref{main}
is established.
\prfe

\section{Proof of Theorem~\ref{lower}}
\setcounter{equation}{0} 

Theorem~\ref{lower} is proven by contradiction. 
There are two main ingredients: The first part (Subsection~\ref{redsec}) is a
general argument to establish that if the estimate in the theorem fails, then 
there exists a non-zero function $g$ such that $D^2 \Hc (f_{0})[g]\leq 0$. The
second (Subsection~\ref{hsec}) is to use the measure-preserving property 
incorporated in our perturbation class $\mathcal{S}_{f_{0}}$ to conclude that 
$g=\{f_{0},h\}$ for some function $h$. This leads to a
contradiction to Lemma~\ref{ks} (Subsection~\ref{contrasec}).

\subsection{Existence of $g\neq 0$ with $D^2 \Hc (f_{0})[g]\leq 0$} 
\label{redsec}

The aim of this subsection is to prove the following result:

\begin{lemma} \label{reduction}
Assume that Theorem \ref{lower} were false. Then there is a
function $g\in L^2(\R^6)$ which is spherically symmetric,
supported in the set $\{E \leq E_0\}$, even in $v$, i.e., $g(x,-v)=g(x,v)$, and 
such that
\be \label{normal}
\frac{1}{8\pi }||\nabla U_{g}||_2^2 =1,
\ee
\be \label{hle}
D^2 \Hc (f_{0})[g] = \frac{1}{2}\int\!\!\!\!\int_{\{f_{0}>0\}}
\Phi_{0}^{\prime \prime }(f_{0})\,g^2\,dv\,dx - 1 \leq 0,
\ee
\be \label{preserve}
\int\!\!\!\!\int \partial_f\Phi(f_{0},L)\,g\,dv\,dx = 0
\ee
for all functions $\Phi$ as specified in the definition of 
$\mathcal{S}_{f_{0}}$.
\end{lemma}

\prf
If Theorem \ref{lower} were false, then for any $n\in \N$  there exists 
$f_{n}\in \mathcal{S}_{f_{0}}$ such that 
\[
d(f_{n},f_{0})<\frac{1}{n}
\]
but
\be \label{hsmall}
\Hc(f_{n})-\Hc(f_{0}) < \frac{1}{8\pi\, n}
||\nabla U_{f_{n}}-\nabla U_0||_2^2,  
\ee
in particular, $f_n \neq f_0$.
We define 
\begin{equation} \label{sigma}
\sigma_{n}:=\frac{1}{\sqrt{8\pi}} ||\nabla U_{f_{n}}-\nabla U_0||_2 ,\
g_{n} := \frac{1}{\sigma _{n}} \left(f_{n}-f_{0}\right)
\end{equation}
so that 
\begin{equation} \label{normaln}
\frac{1}{8\pi }||\nabla U_{g_{n}}||_2^2 =1  
\end{equation}
and $f_n = f_0 + \sigma_n g_n$.
By (\ref{decrel}) and (\ref{sigma}),
\begin{equation}
\sigma _{n}^2\leq d(f_{n},f_{0}) < \frac{1}{n}.  \label{sigmazero}
\end{equation}

\noindent
{\em Bounds on $(g_n)$ and weak limit.}\\
As a first step in the
proof of the lemma we need to establish bounds on the
sequence $(g_n)$ which allow us extract a subsequence that
converges to some $g$ which will be our candidate for the
function asserted in the lemma.
By (\ref{sigma}), (\ref{d}), (\ref{decrel}), and (\ref{hsmall}) we find that
\begin{eqnarray} \label{q''}
&&
\frac{1}{\sigma _{n}^2}\int\!\!\!\!\int
[\Phi_{0}(f_{0}+\sigma_{n}g_{n})-\Phi_{0}(f_{0})+(E-E_{0})\sigma _{n}g_{n}]
\,dv\,dx  - 1 \nonumber \\
&&
\qquad\qquad = \frac{1}{\sigma _{n}^2}\left(d(f_n,f_0)
-\frac{1}{4\pi } \int |\nabla U_{f_{n}}-\nabla U_0|^2 dx \right) \nonumber\\ 
&&
\qquad\qquad = \frac{1}{\sigma _{n}^2} \left(\Hc(f_{n})-\Hc(f_{0})\right)
< \frac{1}{\sigma _{n}^2} 
\frac{1}{8\pi\, n } ||\nabla U_{f_{n}}-\nabla U_0||_2^2 
= \frac{1}{n}.\quad 
\end{eqnarray}
Recalling the estimate (\ref{convest}) which is applicable since
$f_n \in \mathcal{S}_{f_{0}}$ this implies that
\[
1+\frac{1}{n} > \frac{1}{\sigma _{n}^2}\int\!\!\!\!\int [\ldots]\,dv\,dx
\geq C_0 \frac{1}{\sigma _{n}^2} \int\!\!\!\!\int |f_n-f_0|^2\,dv\,dx 
= C_0\, \int\!\!\!\!\int |g_n|^2\,dv\,dx,
\]
which means that the sequence $(g_n)$ is bounded in $L^2(\R^6)$.
Moreover, since the integrand $[\ldots]$ in (\ref{q''}) is non-negative 
we find that
\[
1+\frac{1}{n} > \frac{1}{\sigma _{n}^2}
\int\!\!\!\!\int_{\{E\geq E_0\}} [\ldots]\,dv\,dx
\geq \frac{1}{\sigma _{n}}
\int\!\!\!\!\int_{\{E\geq E_0\}} (E-E_0)\, g_n\,dv\,dx,
\]
where we used the fact that on the set $\{E\geq E_0\}$ the steady
state distribution $f_0$ and hence also $\Phi_0(f_0)$ vanish
while $\Phi_0(f_n)\geq 0$.
By (\ref{sigmazero}),
\be \label{zerooutside1}
\int\!\!\!\!\int_{\{E\geq E_0\}} (E-E_0)\, g_n\,dv\,dx \leq 2 \sigma_n \to 0,\
n\to \infty.
\ee
Now fix any $E_0< E_1 < 0$. Since $\lim_{|x|\to \infty} U_0 (x) = 0$
it follows that $E=E(x,v) > E_1$ for $x$ or $v$ large so that the set
$\{E\leq E_1\}\subset \R^6$
is compact. Hence $(g_n)$ is bounded in $L^1(\{E\leq E_1\})$. 
In addition
%\[
%\int\!\!\!\!\int_{\{E\geq E_1\}} (E-E_0)\, g_n\,dv\,dx 
%\leq \int\!\!\!\!\int_{\{E\geq E_0\}} (E-E_0)\, g_n\,dv\,dx \to 0,
%\]
%and thus
\be \label{zerooutside2}
\int\!\!\!\!\int_{\{E\geq E_1\}} g_n\,dv\,dx 
\leq 
\int\!\!\!\!\int_{\{E\geq E_0\}} \frac{E-E_0}{E_1-E_0}\, g_n\,dv\,dx \to 0;
\ee
notice that $g_n \geq 0$ outside the support of $f_0$, i.e.,
on the set $\{E > E_0\}$. 
Thus we have shown that $(g_n)$ is bounded in $L^1\cap L^2(\R^6)$.
We extract a subsequence, denoted again by $(g_n)$ such that
\[
g_n \rightharpoonup g\ \mbox{weakly in}\ L^2(\R^6).
\]
Since $g_n\geq 0$ on $\{E>E_0\}$ and since (\ref{zerooutside2}) holds
for any $E_0< E_1 < 0$ we conclude that $g$ vanishes a.~e.\ outside
the set $\{E\leq E_0\}$ as desired. Since the functions $g_n$
are spherically symmetric so is $g$.

\smallskip

\noindent
{\em Proof of (\ref{normal}).}\\
In order to pass the week convergence into (\ref{normaln}) 
we need better bounds for the sequence $(g_n)$.
Indeed, we can bound its kinetic energy:
\[
\int\!\!\!\!\int_{\{E\geq E_0\}}
\left(\frac{1}{2} |v|^2 +U_0(x) -E_0\right)\,g_n\,dv\,dx
= \int\!\!\!\!\int_{\{E\geq E_0\}}
\left(E-E_0\right)\,g_n\,dv\,dx \to 0
\]
by (\ref{zerooutside1}) so that 
\[
\int\!\!\!\!\int_{\{E\geq E_0\}}
\frac{1}{2} |v|^2\,g_n\,dv\,dx
\]
is bounded, while
\[
\int\!\!\!\!\int_{\{E\leq E_0\}} \frac{1}{2} |v|^2\,g_n\,dv\,dx 
\leq (E_0 - U_0(0)) \int\!\!\!\!\int_{\{E\leq E_0\}}
g_n\,dv\,dx
\]
is bounded as well; recall that $U_0$ is spherically symmetric 
and radially increasing. Now well known interpolation arguments 
imply that the sequence of induced spatial densities $(\rho_{g_n})$ 
is bounded in $L^1\cap L^{7/5} (\R^3)$, 
cf.\ \cite[Ch.~1, Lemma~5.1]{R3}, so without loss of generality,
\[
\rho_{g_n} \rightharpoonup \rho_g\ \mbox{weakly in}\ L^{7/5}(\R^3).
\]
Let again $E_0<E_1<0$ be arbitrary but fixed and choose $R_1>0$
such that $U_0(R_1)=E_1$ which implies that $E(x,v)\geq E_1$
for $|x|\geq R_1$. Then by (\ref{zerooutside2}),
\[
\int_{\{|x|\geq R_1\}} |\rho_{g_n}|\, dx
\leq
\int\!\!\!\!\int_{\{E\geq E_1\}} g_n\,dv\,dx
\to 0.
\]
The fact that the sequence $(\rho_{g_n})$ remains concentrated in this
way gives the desired compactness:
\[
\nabla U_{g_n} \to \nabla U_g \ \mbox{strongly in}\ L^2(\R^3),
\]
cf.\ \cite[Ch.~2, Lemma~3.2]{R3}. 
Hence we can pass to the limit in (\ref{normaln}) and find that
$g$ satisfies the condition (\ref{normal}) in the lemma.

\smallskip

\noindent
{\em Proof of (\ref{hle}).}\\ 
Since $||g_{n}||_2$ is bounded it follows from (\ref{sigmazero}) that 
\[
||\sigma _{n}g_{n}||_2 \leq C\sigma _{n} \to 0.
\]
Therefore, after extracting again a subsequence,
$\sigma _{n}g_{n}  \to 0$ almost
everywhere.
By Egorov's Theorem there exists for every
$m\in \N$ a measurable 
subset $K_{m} \subset \{E \leq E_0\}$ with the property that 
\[
\vol \left(\{E \leq E_0\}\setminus K_m\right) <\frac{1}{m}
\ \mbox{and}\
\lim_{n \to \infty}\sigma_{n}g_{n}=0 \ 
\mbox{uniformly on}\ K_{m};
\]
note that the set $\{E \leq E_0\}$ has finite measure. In addition we can 
assume that $K_m \subset K_{m+1},\ m\in \N$.
On the set $\{E \leq E_0\}$,
\[
[\Phi_0(f_n) - \Phi_0(f_0) + (E-E_0) \sigma_n g_n] 
 = \frac{1}{2} \Phi_0''(f_0)(\sigma_n g_n)^2
+ \frac{1}{6} \Phi_0'''(f_0+\tau \sigma_n g_n)(\sigma_n g_n)^3
\]
for some $\tau \in [0,1]$. Since both $f_0$ and
$f_0+\sigma_n g_n = f_n$ are non-negative the same is true
for $f_0+\tau \sigma_n g_n$, and we can use the estimate
\[
|\Phi_0''' (f)| = \frac{1}{(1+f)^2} \leq 1,\ f\geq 0
\]
the estimate (\ref{q''}), and the fact that the integrand
$[\ldots]$ in (\ref{q''}) is
non-negative by (\ref{convest}) to conclude that
\beas
\int\!\!\!\!\int_{K_m} \frac{1}{2} \Phi_0''(f_0)\, |g_n|^2\,dv\,dx
&=&
\frac{1}{\sigma_n^2} \int\!\!\!\!\int_{K_m} [\ldots]\,dv\,dx\\
&&
{}- \frac{1}{\sigma_n^2} \int\!\!\!\!\int_{K_m}
\frac{1}{6} \Phi_0'''(f_0+\tau \sigma_n g_n)\,(\sigma_n g_n)^3\,dv\,dx\\
&<&
1+\frac{1}{n} + \sup_{K_m}|\sigma_n g_n| \int\!\!\!\!\int |g_n|^2\,dv\,dx.
\eeas
Taking the limit $n\to \infty $ in this estimate we find that
\[
\int\!\!\!\!\int_{K_m} \frac{1}{2} \Phi_0''(f_0)\, g^2\,dv\,dx \leq 1.
\] 
If we observe the choice of the sets $K_m$, let $m\to \infty$, 
and recall the fact that $g=0$ outside the set $\{E\leq E_0\}$ 
the proof of (\ref{hle}) is complete.

\smallskip

\noindent
{\em Proof of (\ref{preserve}).}\\ 
To prove (\ref{preserve}), the measure-preserving property of the set 
$\mathcal{S}_{f_{0}}$ plays the crucial role. 
Let $\Phi=\Phi(f,L)$ be a function as specified in
the definition of that set respectively in the lemma.
By Taylor expansion with respect to the first argument,
\[
\Phi(f_n,L)-\Phi(f_{0},L)
=\partial _{f}\Phi(f_{0},L)\,\sigma _{n}g_{n}+
\frac{1}{2}\partial_{f}^2\Phi(f_{0}+\tau\sigma_{n}g_{n},L)\,
(\sigma _{n}g_{n})^2
\]
for some $\tau\in[0,1]$. If we integrate this identity
and observe that, since $f_n \in \mathcal{S}_{f_{0}}$,
\[
\int\!\!\!\!\int \Phi(f_n,L)\,dv\,dx=\int\!\!\!\!\int \Phi(f_{0},L)\,dv\,dx,
\]
it follows that
\[
\int\!\!\!\!\int
\partial_{f}\Phi(f_{0},L)\,g_{n}\,dv\,dx 
=
-\frac{1}{2}\sigma_{n}
\int\!\!\!\!\int\partial_f^2 \Phi(f_{0}+\tau \sigma_{n}g_{n},L)\,g_n^2\,dv\,dx
\to 0.
\]
As to the latter limit we note that $\partial_f^2 \Phi$ is bounded,
$(g_n)$ is bounded in $L^2(\R^6)$, and $\sigma _{n}\to 0$. 
On the other hand $\partial_{f}\Phi(f_{0},L)$ is supported
on the compact set $\{E\leq E_0\}$ and hence bounded.
Since $g_{n}\to g$ in $L^2(\R^6)$, the identity (\ref{preserve}) 
follows as $n \to \infty$.

\smallskip
\noindent
{\em Conclusion of the proof of Lemma~\ref{reduction}.}\\
The function $g$ constructed above has all the properties
required in the lemma, except that it need not be even in $v$. Hence
we decompose it into its even and odd parts with respect to $v$: $g=g_\mathrm{even}+g_\mathrm{odd}$. 
We claim that (\ref{normal}), (\ref{hle}), and (\ref{preserve}) 
remain valid for the even part. Since 
\[
\rho_{g}=\int g\, dv=\int g_\mathrm{even}dv=\rho_{g_\mathrm{even}}
\] 
we have $\nabla U_{g_\mathrm{even}}=\nabla U_{g}$, and (\ref{normal})
remains valid. 
Since $\Phi_0^{\prime \prime }(f_0)$ is even in $v$, 
\beas
1
&\geq&
\int\!\!\!\!\int \frac{1}{2} 
\Phi_0''(f_0)\, (g_\mathrm{even}+g_\mathrm{odd})^2\,dv\,dx
=
\int\!\!\!\!\int \frac{1}{2} 
\Phi_0''(f_0)\, \left((g_\mathrm{even})^2+(g_\mathrm{odd})^2\right)\,dv\,dx \\
&\geq&
\int\!\!\!\!\int \frac{1}{2} 
\Phi_0''(f_0)\, (g_\mathrm{even})^2\,dv\,dx,
\eeas
i.e., (\ref{hle}) remains valid. Finally, let $\Phi$ be as in
the definition of the set $\mathcal{S}_{f_{0}}$. Then
$\partial_f \Phi(f_0,L)$ is even in $v$ so that in (\ref{preserve})
the odd part of $g$ drops out, and the assertions of Lemma~\ref{reduction}
hold for $g_\mathrm{even}$.
\prfe

\subsection{Characteristics and $g=\{f_{0},h\}$}\label{hsec}

In this subsection we construct a spherically symmetric
function $h$ such that $g=\{f_{0},h\}$.
To this end we need to introduce variables which are adapted to the
spherical symmetry:
\[
r:=|x|,\ w:= \frac{x\cdot v}{r},\ L:=|x\times v|^2;
\]
$w$ is the radial velocity, and $L$ has already been used above.
Any spherically symmetric function of $x$ and $v$ such as
the desired $h$ can be written
in terms of these variables, so $h=h(r,w,L)$. Then
\[
\{f_{0},h\} 
=
\phi_0'(E) \{E,h\}
= - \phi_0'(E)\,\left[w\, \partial_r +
\left(\frac{L}{r^3}-U_0'(r)\right)\,\partial_w \right]\, h,
\]
and the equation we wish to solve for $h$ reads
\be \label{heqn}
\left[w\, \partial_r +
\left(\frac{L}{r^3}-U_0'(r)\right)\,\partial_w \right]\, h= - \frac{1}{\phi_0'(E)}\,g.
\ee
In order to analyze its characteristic system
\[
\dot r = w,\ \dot w = \frac{L}{r^3}-U_0'(r)
\]
we define for fixed $L>0$ the effective potential
\[
\Psi_L(r):= U_0(r) + \frac{L}{2 r^2}
\]
and observe that in terms of the variables $r,w,L$ the 
conserved particle energy takes the form
\[
E=E(x,v)=E(r,w,L) = \frac{1}{2}w^2 +\Psi_L(r);
\]
$L$ only plays the role of a parameter here since $\dot L = 0$.
We need to analyze the effective potential $\Psi_L$.
The boundary condition for $U_0$ at infinity implies that 
$\lim_{r\to \infty} \Psi_L(r)=0$, and clearly,
$\lim_{r\to 0} \Psi_L(r)=\infty$. Moreover,
\[
U_0'(r)=\frac{4\pi }{r^2}\int_{0}^{r}\rho_{0}(s)\,s^2 ds =:
\frac{m_0(r)}{r^2}>0,\ r>0,
\]
since $U_0$ is increasing, thus $U_0(0) < E_0$, and by (\ref{king}),
$\rho_{0}(0)>0$. Now
\[
\Psi_L'(r) = 0 \ \Leftrightarrow\  \frac{m_0(r)}{r^2}- \frac{L}{r^3} =0
\ \Leftrightarrow\  m_0(r)-\frac{L}{r} = 0,
\]
and since the left hand side of the latter equation
is strictly increasing for $L>0$ with 
$\lim_{r\to \infty}(m_0(r)-L/r) = M >0$ and 
$\lim_{r\to 0}(m_0(r)-L/r) = -\infty$ there exists a unique $r_L >0$
such that
\[
\Psi_L'(r_L)=0,\ \Psi_L'(r)<0\ \mbox{for}\ r<r_L,
\ \Psi_L'(r)>0\ \mbox{for}\ r>r_L.
\]
Moreover, since 
\[
\frac{d}{dr} \left(m_0(r)-\frac{L}{r}\right) 
= 4 \pi r^2 \rho_0(r) + \frac{L}{r^2} >0
\]
the implicit function theorem implies that the mapping
$]0,\infty[ \ni L \mapsto r_L$ is continuously differentiable.
For $r=r_L$,
\be \label{psi''>0}
\Psi''_L (r_L) = - 2\frac{m_0(r)}{r^3} + 3\frac{L}{r^4} 
+ 4 \pi \rho_0(r) = 4 \pi \rho_0(r) + \frac{L}{r^4} > 0.
\ee
The behavior of $\Psi_L$ implies that for any $L>0$ and
$\Psi_L(r_L)<E<0$ there exist two unique 
radii $0<r_{-}(E,L)<r_L<r_{+}(E,L)< \infty$ such that 
\[
\Psi_L(r_\pm(E,L))=E,\ \mbox{and}\ 
\Psi_L(r)<E\ \Leftrightarrow\ r_{-}(E,L)<r<r_{+}(E,L).
\]
Since $\Psi'_L(r)\neq 0$ for $r \neq r_L$ it follows
again by the implicit function theorem that the mapping
$(E,L) \mapsto r_\pm (E,L)$ is $C^1$ on the set
$\{(E,L) \in \R \times ]0,\infty[ \mid \Psi_L(r_L) < E < 0 \}$.

Let $\tau\mapsto (r(\tau),w(\tau),L)$ be a characteristic curve with $L>0$ and
running in the set $\{E\leq E_0\}$, so that
$\Psi_L(r_L)<E\leq E_0 < 0$. Then
\[
r(\tau) \in [r_{-}(E,L),r_{+}(E,L)],\ w(\tau) = \pm \sqrt{2E - 2\Psi_L(r(\tau))}.
\]
By the equation (\ref{heqn}) which we want to solve,
\[
\frac{d}{d\tau} h(r(\tau),w(\tau),L) = -\frac{1}{\phi_0'(E)} g(r(\tau),w(\tau),L),
\]
which we rewrite in terms of the parameter $r$ as
\[
\frac{d}{dr} h(r,w(r),L) = -\frac{1}{\phi_0'(E)\,w(r)} g(r,w(r),L),\ 
w(r) = \pm \sqrt{2E - 2\Psi_L(r)}.
\]
Hence for $(r,w,L) \in \{E\leq E_0\}$ with $L>0$
we define $h$ as follows:
We let $E:= \frac{1}{2} w^2 + \Psi_L(r)$ so that 
$r_{-}(E,L)\leq r \leq r_{+}(E,L)$, and 
\begin{equation}
h(r,w,L) := - \sign w \frac{1}{\phi_0'(E)}\int_{r_{-}(E,L)}^{r}
\frac{g(s,\sqrt{2E-2\Psi_L(s)},L)}{\sqrt{2E-2 \Psi_L(s)}} ds.  \label{hg}
\end{equation}
Outside the set $\{E\leq E_0\}$ we let $h=0$. 
We need to make sure that we can consistently set
$h(r,0,L)=0$, i.e., the integral in the above definition
must vanish for $r=r_+(E,L)$. To this end let
$\Phi$ be as specified in the definition of the class
$\mathcal{S}_{f_{0}}$. We want to apply the change of variables
$(x,v) \mapsto (r,w,L) \mapsto (r,E,L)$ to the integral in (\ref{preserve}).
Now
\be \label{xvtorel}
dx\,dv = 8\pi^2 dr\,dw\,dL = 8\pi^2 
\frac{dr\,dE\,dL}{\sqrt{2E-2\Psi_L(r)}},
\ee
where we note that $g$ is even in $v$ and hence in $w$ 
so that we can restrict the integral to $w>0$. We obtain
the identity
\beas
0
&=&
\int\!\!\!\!\int \partial_f\Phi(f_{0},L)\,g\,dv\,dx
=
8\pi^2 \int_0^\infty \int_0^\infty \int_0^\infty
\partial_f\Phi(f_{0},L)\,g\,dr\,dw\,dL\\
&=&
8\pi^2
\int\!\!\!\!\int_M
\int_{r_{-}(E,L)}^{r_+(E,L)}
\frac{g(r,\sqrt{2E-2\Psi_L(r)},L)}{\sqrt{2E-2 \Psi_L(r)}} dr\,
\partial_f\Phi(\phi_0(E),L)\,dE\,dL,
\eeas
where $M:=\{(E,L)(x,v)|f_0(x,v)>0\}$. 
The class of test functions $\partial_f\Phi(\phi_0(E),L)$ is sufficiently 
large to conclude that for almost all $E$ and $L$,
\[
\int_{r_{-}(E,L)}^{r_{+}(E,L)}\frac{g(r,\sqrt{2E-2\Psi_L(r)},L)}
{\sqrt{2E-2\Psi_L(r)}} dr =0
\]
as desired. 

\subsection{Contradiction to Lemma~\ref{ks}}
\label{contrasec}

As defined above, $h$ need not be smooth or even integrable,
so in order to derive a contradiction to Lemma~\ref{ks}
we need to regularize it. In order to do so it should first
be noted that $h$ as defined in (\ref{hg}) is measurable,
which follows by Fubini's Theorem and the fact that by the change
of variables formula the function
\[
(s,r,E,L) \mapsto \frac{g(s,\sqrt{2 E -2\Psi_L(s)},L)}
{\sqrt{2 E -2\Psi_L(s)}} 
\mathbf{1}_{[\Psi_L(r_L),E_0]}(E)\mathbf{1}_{[r_-(E,L),r_+(E,L)]}(s)
\mathbf{1}_{[0,r]}(s)
\]
is integrable; $r\leq \max\{|x| \mid (x,v) \in \supp f_0\}$.

\smallskip

\noindent{\em The cut-off $h$.}\\
As a first step in regularizing $h$ we define for $m$ large the set
\[
\Omega_{m}:= \left\{(x,v)\in \R^6 \mid
E \leq E_{0}-\frac{1}{m},\ L\geq \frac{1}{m}\right\}.
\]
We want to approximate $h$ by $h\mathbf{1}_{\Omega _{m}}$. 
In order to analyze this approximation the following lemma will be useful:
\begin{lemma} \label{line}
There exists a constant $C_{m}>0$ such that for
$L\geq \frac{1}{m}$ and $\Psi_L(r_L) < E \leq E_0$, 
\[
\int_{r_{-}(E,L)}^{r_{+}(E,L)}\frac{dr}{\sqrt{2 E-2 \Psi_L(r)}} <C_{m}.
\]
\end{lemma}
\prf
We first establish the following auxiliary estimate:
For all $m\in \N$ there exists a constant $\eta_{m}>0$ such that
for all  $L\geq 1/m$, $\Psi_L(r_L)< E\leq E_0$,
and  $r\in[r_-(E,L),r_+(E,L)]$,
\begin{equation} \label{claim}
 \frac{|\Psi'_L(r)|}{\sqrt{\Psi_L(r)-\Psi_L(r_L)}}
\geq \eta _{m}. 
\end{equation}
To see this let $m\in \N$ and $L,E,r$ be as specified.
Then
\[
E_0 \geq E \geq \Psi_L(r) = U_0(r) + \frac{L}{2 r^2} 
\geq U_0(0) + \frac{1}{2 m r^2}
\]
and hence $r\geq (2m(E_0 - U_0(0))^{-1/2}$. 
Let $R:=\max\{|x| \mid (x,v)\in \supp f_0\}$, i.e., $U_0(R)=E_0$.
Then
\[
L \leq 2 r^2 \left(E_0 - U_0(r)\right) 
\leq 2 R^2 \left(E_0 - U_0(0)\right).
\]
Hence if we assume that (\ref{claim}) were false, 
we can find a sequence $(r_{n},L_{n})\to (\bar{r},\bar{L})$ 
in the set 
$[(2m(E_0 - U_0(0))^{-1/2},R]\times[1/m, 2 R^2 \left(E_0 - U_0(0)\right)]$
such that 
\[
\lim_{n\to \infty }\frac{\Psi'_{L_n}(r_n)}
{\sqrt{\Psi_{L_{n}}(r_{n}))- \Psi_{L_{n}}(r_{L_{n}})}}=0.
\]
If $\bar{r}\neq r_{\bar{L}}$ it follows that
$\Psi_{\bar L} (\bar r) > \Psi_{\bar L} (r_{\bar L})$ and
$\Psi'_{\bar L}(\bar r) = 0$ which is a contradiction
to the uniqueness of the minimizer $r_{\bar L}$ of 
$\Psi_{\bar L}$. So assume that $\bar{r} = r_{\bar{L}}$.
Now recall that $\Psi'_{L_n}(r_{L_n})=0$.
By Taylor expansion at $r=r_{L_n}$ we find intermediate values
$\theta_n$ and $\tau_n$ between $r_n$ and $r_{L_n}$ such that
\[
\frac{|\Psi'_{L_{n}}(r_{n})|}
{\sqrt{\Psi_{L_{n}}(r_{n}) - \Psi_{L_{n}}(r_{L_n})}}
=
\frac{|\Psi''_{L_{n}}(\theta_{n})\,(r_{n}-r_{L_{n}})|}
{\sqrt{\frac{1}{2} \Psi''_{L_{n}}(\tau_{n})\,(r_{n}-r_{L_{n}})^2}}
\to
\sqrt{2 |\Psi''_{\bar L}(r_{\bar L})|} \neq 0,\ n\to \infty,
\]
where we recall (\ref{psi''>0}).
This contradiction completes the proof of (\ref{claim}).

To complete the proof of the lemma we split the integral as
\[
\int_{r_{-}(E,L)}^{r_{+}(E,L)}\frac{dr}{\sqrt{2E-2\Psi_L(r)}}
=\int_{r_{-}(E,L)}^{r_L}\ldots + \int_{r_L}^{r_{+}(E,L)}\ldots 
=: I_1 + I_2.
\]
In the first term, we make a change of variables 
$u=\sqrt{\Psi_L(r)-\Psi_L(r_L)}$ so that
$\frac{du}{dr}=\frac{1}{2 u}\Psi_L'(r) <0$ on $[r_{-}(E,L),r_L[$.
By (\ref{claim}), 
\begin{eqnarray*}
I_1
&=&
\int_{\sqrt{E-\Psi_L(r_L)}}^{0}
\frac{1}{\sqrt{2(E-\Psi_L(r_L)-u^2)}}\frac{dr}{du}du \\
&\leq&
\frac{\sqrt{2}}{\eta _{m}}
\int_{0}^{\sqrt{E-\Psi_L(r_L)}}
\frac{du}{\sqrt{E-\Psi_L(r_L)-u^2}}
=
\frac{\sqrt{2}}{\eta _{m}}
\int_{0}^{1}\frac{ds}{\sqrt{1-s^2}}<\infty 
\end{eqnarray*}
by a further change of variables $u=\sqrt{E-\Psi_L(r_L)} s$. 
The same type of estimate holds for the second part
$I_2$ of the integral under investigation, and the proof of the
lemma is complete.
\prfe

We now show that the cut-off function $h\mathbf{1}_{\Omega _{m}}$
is square integrable and solves the equation
$\{f_{0},h\mathbf{1}_{\Omega _{m}}\} =g\mathbf{1}_{\Omega _{m}}$
in the sense of distributions, more precisely:
\begin{lemma} \label{dis}
For any $m\in \N$ large, 
$h\mathbf{1}_{\Omega _{m}}\in L^2(\R^6)$, and 
for any spherically symmetric test function 
$\psi=\psi(r,w,L) \in C^1([0,\infty[\times \R \times [0,\infty[)$,
\[
\int_{E\leq
E_{0}}\{f_{0},\psi \}h\mathbf{1}_{\Omega _{m}}=-
\int_{E\leq E_{0}}g\mathbf{1}_{\Omega _{m}}\psi.
\]
\end{lemma}
\prf
We first prove that $h\mathbf{1}_{\Omega _{m}}\in L^2(\R^6)$. 
Since the integrand is even in $v$ we can apply
the change of variables (\ref{xvtorel}):
\beas
&&
\int\!\!\!\!\int \mathbf{1}_{\Omega _{m}} h^2 dv\,dx\\
&&
\qquad
= 8 \pi^2 \int\!\!\!\!\int_{S_m} 
\int_{r_{-}(E,L)}^{r_{+}(E,L)}h^2(r,\sqrt{2E-2\Psi_L(r)},L)
\frac{dr}{\sqrt{2E-2\Psi_L(r)}}\,dE\,dL,
\eeas
where 
\be \label{sm}
S_m:= 
\left\{(E,L)=(E,L)(x,v)
\mid (x,v)\in \supp f_0,\ E\leq E_0-\frac{1}{m},\  L\geq \frac{1}{m}\right\}.
\ee
Let $(E,L) \in S_m$. In the estimates below we write
$w(r)=\sqrt{2E-2\Psi_L(r)}$ and $r_\pm(E,L) = r_\pm$
for brevity. Then by the definition
(\ref{hg}) of $h$ and the fact that $1/|\phi_0'(E)| \leq 1$ for $E\leq E_0$,
\begin{eqnarray*}
\int_{r_{-}}^{r_{+}}h^2(r,w(r),L) \frac{dr}{w(r)}
&=&
\int_{r_{-}}^{r_{+}}
\left[\frac{1}{\phi_0'(E)} \int_{r_{-}}^{r}
g(s,w(s),L) \frac{ds}{w(s)}\right]^2
\frac{dr}{w(r)} \\
&\leq&
\int_{r_{-}}^{r_{+}}
\left[\int_{r_{-}}^{r_{+}}
g(s,w(s),L) \frac{ds}{w(s)}\right]^2
\frac{dr}{w(r)} \\
&\leq&
\left(\int_{r_{-}}^{r_{+}}\frac{dr}{w(r)}\right)^2
\int_{r_{-}}^{r_{+}}
g^2(r,w(r),L) \frac{dr}{w(r)}\\
&\leq&
C_m^2
\int_{r_{-}}^{r_{+}}
g^2(r,w(r),L) \frac{dr}{w(r)};
\end{eqnarray*}%
in the last two estimates
we used the Cauchy-Schwarz inequality and Lemma~\ref{line}.
A further integration with respect to $E$ and $L$ and the change
of variables  $(r,E,L)\mapsto (x,v)$ shows that 
$||h\mathbf{1}_{\Omega _{m}}||_2$ is bounded in terms of $C_m$ and
$||g||_2$.

Now let $\psi=\psi (r,w,L)$ be a test function 
as specified in the lemma. Along characteristic curves
of (\ref{heqn}), which as before we parameterize by $r$
distinguishing between $w>0$ and $w<0$,
a simple computation shows that
\[
\{E,\psi\}= -\sign w \sqrt{2 E- 2 \Psi_L(r)}
 \frac{d}{dr}[\psi (r,\sign w \sqrt{2 E- 2 \Psi_L(r)},L)].
\]
By the change of variables used repeatedly above, using the
abbreviation $w(r):=\sqrt{2 E- 2 \Psi_L(r)}$, and recalling
the definition (\ref{sm}) we find that 
\begin{eqnarray*}
&&
\int\!\!\!\!\int \{f_{0},\psi \}h\mathbf{1}_{\Omega _{m}}dv\,dx
=\int_{\{w>0\}} \ldots + \int_{\{w<0\}} \ldots\\
&&
= -\int\!\!\!\!\int_{S_m} \phi_0'(E)
\int_{r_{-}}^{r_{+}} w(r)
\frac{d}{dr}[\psi (r,w(r),L)]\,
h(r,w(r),L) \frac{dr}{w(r)} dE\,dL\\
&& {}-\int\!\!\!\!\int_{S_m} \phi_0'(E)
\int_{r_{-}}^{r_{+}} (-w(r))
\frac{d}{dr}[\psi (r,-w(r),L)]\,
h(r,-w(r),L) \frac{dr}{w(r)} dE\,dL\\
&&
= \int\!\!\!\!\int_{S_m} \phi_0'(E)
\int_{r_{-}}^{r_{+}} \psi (r,w(r),L)\, 
\left(-\frac{g(r,w(r),L)}{\phi_0'(E)\, w(r)}\right)\,dr\, dE\,dL\\
&&
{}- \int\!\!\!\!\int_{S_m} \phi_0'(E)
\int_{r_{-}}^{r_{+}} \psi (r,-w(r),L)\, 
\frac{g(r,-w(r),L)}{\phi_0'(E)\, w(r)}dr\, dE\,dL\\
&&
= - \int\!\!\!\!\int \psi g \mathbf{1}_{\Omega _{m}}dv\,dx;
\end{eqnarray*}
notice that $h(r,\pm w(r),L)=0$ for $r=r_\pm(E,L)$, which 
together with the definition (\ref{hg}) of $h$ along the
characteristics was essential in the integration by parts
above, and also that $g$ is even in $v$ respectively $w$.
The proof of Lemma~\ref{dis} is now complete.
\prfe

\smallskip

\noindent
{\em Regularization of $h\mathbf{1}_{\Omega _{m}}$}.\\
The function $h\mathbf{1}_{\Omega _{m}}$ is not smooth,
hence Lemma~\ref{ks} cannot be applied to it, and therefore
we smooth it. For fixed $m\in \N$ the function $h\mathbf{1}_{\Omega _{m}}$
is, as a function of $r,w,L$, supported in a set of the form
\be \label{supphm}
Q_m := [R_0,R_1]\times [-W_0,W_0]\times [L_0,L_1]
\ee 
with $0<R_0<R_1$, $W_0>0$, and $0<L_0<L_1$; it will be important
that its support stays away both from $r=0$ and $L=0$.
Let $\zeta \in C_c^\infty (\R^3)$ be even in all three
variables, i.e., 
$\zeta(z_1,z_2,z_3) = \zeta(|z_1|,|z_2|,|z_3|)$, $\zeta \geq 0$,
$\int \zeta =1$, and define $\zeta_n :=n^3 \zeta(n\cdot)$ for $n\in \N$.
For $n$ sufficiently large we define 
\[
h_n (r,w,L)
=\int_0^\infty \int_{-\infty}^\infty \int_0^\infty
(h\mathbf{1}_{\Omega _{m}})(\bar r,\bar w,\bar L)\,
\zeta _{n}(r-\bar r,w-\bar w,L-\bar L)\,d\bar L\,d\bar w\,d\bar r.
\]
Since $\Omega_m$ is strictly inside the set
$\{E\leq E_0\}$ with a positive distance from its boundary,
$h_n \in C^\infty_c(]0,\infty[\times \R \times ]0,\infty[)$, 
and without loss of generality we can assume that 
$\supp h_n \subset Q_m \cap \{E < E_0\}$.
Since $h$ is odd in $w$ respectively $v$ so is $h_n$, and clearly,
$h_n  \to h \mathbf{1}_{\Omega _{m}}$ in $L^2\cap L^{1}$.
The crucial step is to show that
\be \label{hnbracketlimit}
\lim_{n \to \infty }\{f_{0},h_n\}=g\mathbf{1}_{\Omega _{m}}
\ee
in $L^2$. To this end we fix $(r,w,L)$, write
$\bar E :=E(\bar r,\bar w, \bar L)$ and 
$\int:=\int_0^\infty \int_{-\infty}^\infty \int_0^\infty$ for brevity, 
and split the convolution integral into three parts:
\beas
&&
\{f_{0},h_n\} =
-\phi_0'(E)
\left[w\partial_r +\left(\frac{L}{r^3}-U_0'(r)\right)\partial_w\right]\,h_n \\
&&
\quad = -\int
(h\mathbf{1}_{\Omega_{m}})(\bar r,\bar w,\bar L)(\phi_0'(E)-\phi_0'(\bar E))
\left[w\partial_r +\left(\frac{L}{r^3}-U_0'(r)\right)\partial_w\right]\\
&&
\quad \qquad \qquad \qquad \qquad \qquad \qquad \qquad \qquad \qquad
\zeta _{n}(r-\bar r,w-\bar w,L-\bar L)\,d\bar L\,d\bar w\,d\bar r\\
&&
\quad {} +
\int
(h\mathbf{1}_{\Omega _{m}})(\bar r,\bar w,\bar L) \phi_0'(\bar E)\,
\left[\bar w\partial_{\bar r} +\left(\frac{\bar L}{\bar r^3}-
U_0'(\bar r)\right)\partial_{\bar w}\right]\\
&&
\quad \qquad \qquad \qquad \qquad \qquad \qquad \qquad \qquad \qquad
\zeta _{n}(r-\bar r,w-\bar w,L-\bar L)\,d\bar L\,d\bar w\,d\bar r\\
&&
\quad {} -
\int
(h\mathbf{1}_{\Omega _{m}})(\bar r,\bar w,\bar L)\phi_0'(\bar E)\,
\left[(\bar w-w)\partial_{\bar r} +
\left(\frac{\bar L}{\bar r^3}-\frac{L}{r^3} -
U_0'(\bar r) + U_0'(r)\right)\partial_{\bar w}\right]\\
&&
\quad \qquad \qquad \qquad \qquad \qquad \qquad \qquad \qquad \qquad
\zeta _{n}(r-\bar r,w-\bar w,L-\bar L)\,d\bar L\,d\bar w\,d\bar r\\
&&
\quad =: - I_1 + I_2 - I_3 .
\eeas
By Lemma \ref{dis}, 
\[
I_2=
\int g\mathbf{1}_{\Omega _{m}}(\bar r,\bar w,\bar L)
\zeta _{n}(r-\bar r,w-\bar w,L-\bar L)\,d\bar L\,d\bar w\,d\bar r
\to g\mathbf{1}_{\Omega _{m}}\ \mbox{in}\ L^2.
\] 
We show that $I_1$ and $I_3$ tend to zero in $L^2$. To do so we
change the integration variables into 
$\tilde r = n (r-\bar r)$, $\tilde w = n (w-\bar w)$,
$\tilde L = n (L-\bar L)$ so that
$\bar r = r-\tilde r/n$, $\bar w = r-\tilde w/n$,
$\bar L = r-\tilde L/n$, 
$d\bar L\,d\bar w\,d\bar r = n^{-3}d\tilde L\,d\tilde w\,d\tilde r$,
and $\partial_r \zeta_n(r-\bar r,w-\bar w,L-\bar L)=
n^4 \partial_{\tilde r} \zeta (\tilde r, \tilde w, \tilde L)$
with analogous formulas for the other derivatives. Thus
\beas
I_1
&=&
\int (h\mathbf{1}_{\Omega_{m}})(\bar r,\bar w,\bar L)
\frac{\phi_0'(E)-\phi_0'(\bar E)}{1/n}\\
&&
\quad
\left[w\partial_{\tilde r} +\left(\frac{L}{r^3}-U_0'(r)\right)
\partial_{\tilde w}\right]\,
\zeta (\tilde r, \tilde w, \tilde L)\,d\tilde L\,d\tilde w\,d\tilde r,\\
I_3
&=&
\int (h\mathbf{1}_{\Omega_{m}})(\bar r,\bar w,\bar L)\, \phi_0'(\bar E)\\
&&
\quad
\left[-\tilde w\partial_{\tilde r} +
\left(- \frac{\tilde L}{\bar r^3} + L \frac{\bar r ^{-3} - r^{-3}}{1/n}
+ \frac{U_0'(\bar r)-U_0'(r)}{1/n}\right)
\partial_{\tilde w}\right]\,
\zeta (\tilde r, \tilde w, \tilde L)\,d\tilde L\,d\tilde w\,d\tilde r
\eeas
Now
\[
\frac{\phi_0'(E)-\phi_0'(\bar E)}{1/n} \to
-\phi_0''(E)\,\left( w \tilde w + \frac{\tilde L}{2r^2} -
\frac{\tilde r L}{r^3} + \tilde r U_0'(r)\right),
\]
\[
\frac{\bar r ^{-3} - r^{-3}}{1/n} \to
\frac{3\tilde r}{r^4},\qquad
\frac{U_0'(\bar r)-U_0'(r)}{1/n} \to
- U_0''(r)\, \tilde r
\]
for $n \to \infty$, and all these limits are uniform with respect to 
$(r,w,L) \in Q_m$ defined in (\ref{supphm}) and with respect
to $(\tilde r,\tilde w,\tilde L)\in \supp \zeta$.
Hence
\beas
&&
I_1 \to
- (h\mathbf{1}_{\Omega _{m}})(r,w,L)\phi_0''(E)\\
&&
\quad \int\left[ w \tilde w + \frac{\tilde L}{2r^2} -
\frac{\tilde r L}{r^3} + \tilde r U_0'(r)\right]
\left[w\partial_{\tilde r} +\left(\frac{L}{r^3}-U_0'(r)\right)
\partial_{\tilde w}\right]
\zeta (\tilde r, \tilde w, \tilde L)\,d\tilde L\,d\tilde w\,d\tilde r\\
&&
= 0,\\
&&
I_3 \to
(h\mathbf{1}_{\Omega _{m}})(r,w,L)\phi_0'(E)\\
&&
\quad \int\left[ -\tilde w \partial_{\tilde r} + 
\left(-\frac{\tilde L}{\tilde r^3} +
\frac{3 L \tilde r}{r^4} - U_0''(r) \tilde r\right)\partial_{\tilde w}
\right]\,
\zeta (\tilde r, \tilde w, \tilde L)\,d\tilde L\,d\tilde w\,d\tilde r \\
&&
=0
\eeas
where both limits are in $L^2$ and the zeroes result from integration
by parts with respect to $\tilde r$ and $\tilde w$ and an exact cancellation. 
The assertion (\ref{hnbracketlimit}) is now established.

\smallskip

\noindent
{\em Finally: The desired contradiction.}\\
The functions $h_n$
have all the properties required in Lemma~\ref{ks}, and we obtain
the estimate 
\[
D^2 \Hc (f_{0})[\{f_{0},h_n\}]\geq -\frac{1}{2}
\int_{\{E<E_0\}} \phi_0'(E)\,
\frac{1}{r}\,U_0'(r)\, |h_n|^2\, dv\,dx.
\]
By (\ref{hnbracketlimit})
and the fact that $h_n$ and $g \mathbf{1}_{\Omega_{m}}$ are supported
in a common compact set,
$\nabla U_{\{f_0,h_n\}} \to \nabla U_{g\mathbf{1}_{\Omega _{m}}}$
in $L^2$ as $n\to \infty$. Since $\frac{1}{r}U_0'(r)$ is bounded on
$[0,\infty[$ we obtain the estimate
\[
D^2 \Hc (f_{0})[g\mathbf{1}_{\Omega _{m}}]
\geq -\frac{1}{2} \int_{\{E<E_0\}} \phi_0'(E)
\frac{1}{r}U_0'(r)\, h^2 \mathbf{1}_{\Omega _{m}}\, dv\,dx.
\]
Since $\lim_{m \to \infty }g\mathbf{1}_{\Omega _{m}}=g$ in $L^{1}\cap
L^2(\R^6)$, there exists $m_{0}$ sufficiently large such that 
$g\mathbf{1}_{\Omega_{m_{0}}}\neq 0$ and by Lemma~\ref{dis},
$h\mathbf{1}_{\Omega _{m_{0}}}\neq 0$.
For all $m\geq m_{0}$, $\Omega _{m_{0}}\subset \Omega _{m}$ and 
$h^2 \mathbf{1}_{\Omega _{m_{0}}}\leq h^2\mathbf{1}_{\Omega _{m}}$. 
Since $\phi_0'(E)<0$ and $\frac{1}{r}U_0'(r)>0$ we have for all 
$m\geq m_{0}$, 
\[
D^2 \Hc (f_{0})[g\mathbf{1}_{\Omega _{m}}]
\geq -\frac{1}{2}\int_{\{E<E_0\}} \phi_0'(E)
\frac{1}{r}U_0'(r)\, h^2 \mathbf{1}_{\Omega _{m_0}}\, dv\,dx =: C >0.
\]
Hence $m \to \infty$ leads to a contradiction
to Lemma~\ref{reduction}, and the proof of Theorem~\ref{lower}
is complete.

\section{Appendix: Proof of Lemma~\ref{ks}}
\setcounter{equation}{0}

Let
\[
U_{h}(x) := \int\!\!\!\!\int \{f_{0},h\} dv\frac{dy}{|x-y|}
\]
be the potential induced by $-\{f_{0},h\}$, which is spherically symmetric.
A short computation using the definition (\ref{lie}) of the Lie-Poisson bracket
shows that
\[
\int \{f_{0},h\} dv = \nabla _{x}\cdot \int v\, \phi_0'(E)\, h(x,v)\,dv.
\]
By the spherical symmetry of $U_h$ and $h$, 
\beas
U_{h}^{\prime }(r)
&=&
\frac{1}{r^2}\int_{|x|\leq r} \nabla _{x}\cdot \int v\, \phi_0'(E)\, h(x,v)\,dv\\
&=&
\frac{1}{r^2}\int_{|x|= r} \int \phi_0'(E)\, h(x,v) v\cdot\frac{x}{r} dv\, d\omega(x)
= 4 \pi \int w \phi_0'(E)\, h(x,v)\, dv.
\eeas
Therefore, by the Cauchy-Schwarz inequality,  
\[
\frac{1}{8\pi }\int |\nabla _{x}U_{h}|^2dx
\leq
2\pi \int \left[ -\int w^2 \phi_0'(E)\,dv\right]\,
\left[ -\int \phi_0'(E)\, h^2 dv\right]\, dx.
\]
Since 
\[
w^2 \phi_0'(E) = w \frac{d}{dw}\phi_0\left(\frac{1}{2} w^2 +\frac{L}{2 r^2}+U_0(r)\right)
\]
an integration by parts with respect to $w$ yields 
\[
-\int w^2 \phi_0'(E)\,dv = \frac{\pi}{r^2}\int_0^\infty\int_{-\infty}^\infty
\phi_0\left(\frac{1}{2} w^2 +\frac{L}{2 r^2}+U_0(r)\right)\, dw\, dL = \rho_0(r).
\]
Hence 
\[
D^2 \Hc (f_{0})[\{f_{0},h\}] 
\geq 
-\frac{1}{2} \int\!\!\!\!\int
\phi_0' (E)\,\left(|\{E,h\}|^2 - 4\pi \rho_{0} h^2\right)\,dv\,dx.
\]
Since $h$ is odd in $v$ respectively $w$ the function
\[
\mu(r,w,L)
:= \frac{1}{r w} h(r,w,L)
\]
is smooth away from $r=0$; in passing we notice that the functions $h_n$
to which we applied Lemma~\ref{ks} in the proof of Theorem~\ref{lower}
have support bounded away from $r=0$, but this is not necessary for Lemma~\ref{ks}.
Since $h=r w \mu$,  
\[
\{E,h\} = rw \{E,\mu \} + \mu \{E,rw\},
\]
and hence
\begin{eqnarray*}
|\{E,h\}|^2 &=&(rw)^2|\{E,\mu \}|^2+rw\{E,rw\}\{E,\mu
^2\}+\mu ^2|\{E,rw\}|^2 \\
&=&
(rw)^2|\{E,\mu \}|^2 + \{E,\mu ^2rw\{E,rw\}\}-\mu^2rw\{E,\{E,rw\}\}.
\end{eqnarray*}
The first term is as claimed in Lemma~\ref{ks}. 
The second term leads to 
$\{f_{0},\mu ^2rw\{E,rw\}\}$ whose integral with respect to $x$ and $v$
vanishes; if we cut a small ball of radius $\epsilon$
about $x=0$ from the $x$-integral
then the surface integral appearing after the integration by parts
with respect to $x$ vanishes for $\epsilon \to 0$ since $r \mu^2 \leq C/r$. 
By the Poisson equation, 
\[
\{E,\{E,rw\}\}=-2 w U_0' - w (r U_0')'=- r w\, \left(4\pi \rho_{0} + \frac{1}{r}U_0'\right).
\]
Hence the third term above becomes
$ 4\pi \rho _{0}h^2 + h^2 \frac{1}{r}U_0'$, and Lemma~\ref{ks} is proven.
\prfe

\noindent
{\bf Acknowledgment.}
The research of the first author 
is supported in part by an NSF grant. This article is dedicated
to the memory of Xudong Liu.


\begin{thebibliography}{SDLP}

\bibitem[An]{An} 
Antonov, V. A. :
Remarks on the problem of stability in stellar
dynamics. 
{\em Soviet Astr, AJ.}, {\bf 4}, 859--867 (1961)

\bibitem[BT]{BT} 
Binney, J., Tremaine, S.: 
{\em Galactic Dynamics}. 
Princeton: Princeton University Press 1987

\bibitem[BG]{BG} 
Burchard, A., Guo, Y.: 
Compactness via symmetrization. 
{\em J.\ Funct.\ Anal.} {\bf 214}, 40--73 (2004)

\bibitem[DSS]{DSS} 
Dolbeault, J., S\'{a}nchez, \'{O}., Soler, J.: 
Asymptotic behaviour for the Vlasov-Poisson system in the stellar-dynamics case.
{\em Arch.\ Ration.\ Mech.\ Anal.}\
{\bf 171}, 301--327 (2004)

\bibitem[FP]{FP} 
Fridman, A., Polyachenko, V.:
{\em Physics of Gravitating Systems I}.
New York: Springer-Verlag 1984

\bibitem[G1]{G1} 
Guo, Y.: 
Variational method in polytropic galaxies.
{\em Arch.\ Ration.\ Mech.\ Anal.}\
{\bf 150}, 209--224 (1999)

\bibitem[G2]{G2} 
Guo, Y.: 
On the generalized Antonov's stability criterion.
{\em Contemp.\ Math.}\ 
{\bf 263}, 85--107 (2000)

\bibitem[GR1]{GR1} 
Guo, Y., Rein, G.:
Stable steady states in stellar dynamics.
{\em Arch.\ Ration.\ Mech.\ Anal.}\
{\bf 147}, 225--243 (1999)

\bibitem[GR2]{GR2} 
Guo, Y., Rein, G.:
Existence and stability of Camm type steady states in galactic 
dynamics.
{\em Indiana Univ.\ Math.\ J.}\
{\bf 48}, 1237--1255 (1999)

\bibitem[GR3]{GR3} 
Guo, Y., Rein, G.: 
Isotropic steady states in galactic dynamics.
{\em  Comm.\ Math.\ Phys.}\ 
{\bf 219}, 607--629 (2001)

\bibitem[GR4]{GR4} 
Guo, Y., Rein, G.: 
Isotropic steady states in stellar dynamics revisited. 
ArXiv preprint math-ph/0010031

\bibitem[H]{H} 
Had\v{z}i\'{c}, M.:
{\em Compactness and stability of some systems of nonlinear PDEs in galactic dynamics}. 
Diploma thesis, University of Vienna 2005

\bibitem[KS]{KS} 
Kandrup, H., Sygnet, J. ~F.: 
A simple proof of dynamical stability for a class of spherical clusters. 
{\em Astrophys.~J.}\
{\bf 298}, 27--33 (1985)

\bibitem[LMR]{LMR} 
Lemou, M., Mehats, F., Raphael, P.:
On the orbital stability of the ground states and the 
singularity formation for the gravitational Vlasov-Poisson system. 
Preprint 2005

\bibitem[LP]{LP}
Lions, P.-L., Perthame, B.:
Propagation of moments and regularity for the 3-dimensional
Vlasov-Poisson system.
{\em Invent.\ Math.}\
{\bf 105}, 415--430 (1991)

\bibitem[Pf]{Pf}
Pfaffelmoser, K.:
Global classical solutions of the Vlasov-Poisson system in three
dimensions for general initial data.
{\em J.\ Differential Equations}\ 
{\bf 95}, 281--303 (1992)

\bibitem[R1]{R1} 
Rein, G.:
Reduction and a concentration-compactness principle
for energy-Casimir functionals.
{\em SIAM J.\ Math.\ Anal.}\
{\bf 33}, 896--912 (2002)

\bibitem[R2]{R2}
Rein, G.:
Nonlinear stability of Newtonian galaxies and stars from a 
mathematical perspective. 
{\em In: Nonlinear Dynamics in Astronomy and Physics,
Annals of the New York Academy of 
Sciences}
{\bf 1045}, 103--119 (2005)

\bibitem[R3]{R3} 
Rein, G.: 
Collisionless Kinetic Equations from Astrophysics---The Vlasov-Poisson system. 
Preprint 2005

\bibitem[RG]{RG} 
Rein, G., Guo, Y.: 
Stable models of elliptical galaxies. 
{\em Monthly Not.\ Royal Astr.\ Soc.}\
{\bf 344}, 1396--1406 (2003)

\bibitem[RR]{RR}
Rein, G., Rendall, A.:
Compact support of spherically symmetric equilibria in 
non-relativistic and relativistic galactic dynamics.
{\em Math.\ Proc.\ Camb.\ Phil.\ Soc.}\
{\bf 128}, 363--380 (2000)

\bibitem[SS]{SS}  
S\'{a}nchez, \'{O}., Soler, J.: 
Orbital stability for polytropic galaxies. Preprint HYKE 2004-043

\bibitem[Sch]{Sch} 
Schaeffer, J.: 
Global existence of smooth solutions to the Vlasov-Poisson system
in three dimensions
{\em Comm.\ Partial Differential Equations}  
{\bf 16}, 1313--1335 (1991)

\bibitem[SDLP]{SDLP}
Sygnet, J.~F., Des Forets, G., Lachieze-Rey, M., Pellat, R.:
Stability of gravitational systems and gravothermal catastrophe
in astrophysics.
{\em Astrophys.~J.}\
{\bf 276}, 737--745 (1984)

\bibitem[Wa]{Wa} 
Wan, Y-H.: 
On nonlinear stability of isotropic models in stellar dynamics.
{\em Arch.\ Ration.\ Mech.\ Anal.}\
{\bf 147},  245--268 (1999)

%\bibitem[Wa2]{Wa2}
%Wan, Y-H.: 
%Nonlinear stability of spherical systems in galactic dynamics. 
%Preprint, 2000

\bibitem[Wo1]{Wo1}
Wolansky, G.: 
On nonlinear stability of polytropic galaxies.
{\em Ann.\ Inst.\ H.\ Poincar\'e Anal.\ Non Lin\'eaire}
{\bf 16}, 15--48 (1999)

\bibitem[Wo2]{Wo2}
Wolansky, G.: 
Static solutions of the Vlasov-Einstein system.
{\em Arch.\ Ration.\ Mech.\ Anal.}\
{\bf 156}, 205--230 (2001)

\end{thebibliography}
\end{document}